\documentclass{article}

\usepackage{arxiv}

\usepackage[utf8]{inputenc} 
\usepackage[T1]{fontenc}    
\usepackage{hyperref}       
\usepackage{url}            
\usepackage{booktabs}       
\usepackage{amsfonts}       
\usepackage{nicefrac}       
\usepackage{microtype}      
\usepackage{lipsum}
\usepackage{color}
\usepackage{algorithm}
\usepackage{algorithmic}
\usepackage{amsmath,amsfonts,amssymb}
\usepackage{subfigure}
\usepackage{lineno}
\usepackage{graphicx}
\usepackage{bm}
\usepackage{subfigure}

\title{Extracting informative vortical structures of turbulent wake-extreme vortex gust interactions with machine learning}

\author{
Ryo Koshikawa and Kai Fukami$^{*}$
\\
\\
Department of Aerospace Engineering, Graduate School of Engineering, Tohoku University, Sendai, 980-8579, Japan\\
$^{*}$Corresponding author: kfukami1@tohoku.ac.jp
}

\begin{document}
\maketitle

\begin{abstract}

\vspace{-2mm}
This study considers extracting causally important vortical structures from the extreme vortex gust-airfoil interaction at a chord-based Reynolds number of $5000$.
This extraction is achieved by decomposing a given vortical flow snapshot into its informative and residual components based on the contribution to an arbitrary future target variable with convolutional information-theoretic learning.
For the current vortex-airfoil interactions that exhibit transient and multiscale flow characteristics, we first examine the important vortical structures with respect to a future lift coefficient.
While the vortex cores are primarily highlighted before vortex impingement, the emerging shear layers are additionally captured after the massive separation, which is evident from a comparison to an instantaneous force-element analysis.
We further take the scale-dependent energy transfer as a future variable of interest to examine its impact on the extracted informative structures compared to the lift-associated structures.
They are distinct from the lift-based structures in the early stage of the gust encounter yet become similar after impingement, revealing an analogy between informative structures across different transient aerodynamic mechanisms.
The present data-driven approach selectively extracts the specific important flow structures responsible for the physics of interest, which can support studying a range of transient aerodynamic flows from the causal, data-driven perspective.

\end{abstract}

\section{Introduction}
\label{sec:intro}
\vspace{-2mm}
{Small- and medium-sized aircraft encounter a variety of transient gust forms during their flight maneuver.
To achieve stable flight even under severe airspace, it is critical to identify key vortical structures with respect to future aerodynamic events~\cite{fukami2023grasping,jones2022physics}.  
A comprehensive understanding of such relationships between flows and an event of interest over time has been tackled with a concept of a cause-and-effect mechanism~\cite{lozano2022information}.
We discuss in this study that it is possible to extract transient, target-event-oriented important structures of turbulent wake-extreme vortex gust interactions from a given flow snapshot using information-theoretic machine learning.}

{An aerodynamic scenario where the gust velocity exceeds the cruising velocity of the vehicle, referred to as {\it extreme aerodynamics}, is of interest~\cite{jones2022physics,taira2026extreme,fukami2023grasping}.
Two-dimensional extreme vortex gust-airfoil interactions have been examined by exploring nonlinear machine-learning-based data compression~\cite{fukami2023grasping}, flow control strategies~\cite{fukami2024data}, and the aerodynamic influence of wing geometry~\cite{lopez2026effect}.
There exist several works focusing on vortex gust encounters with a finite wing, examining the wing tip effect~\cite{barnes2018clockwise,barnes2018counterclockwise,odaka2026extremeb}.
Furthermore, the topological similarity of dominant vortical structures across the Reynolds number has been examined~\cite{odaka2026vorticala}.
In addition to these efforts with numerical simulations, experimental endeavors of flow under such strong disturbance have been carried out~\cite{biler2021experimental,andreu2020effect}.}

{While the interaction with extremely strong gusts generates complex vortical motions, the specific structures that warrant attention vary significantly depending on the aerodynamic event of interest. 
To reveal the underlying physics from flow fields prepared with numerical simulations or experiments, a range of approaches has been considered.
For instance, to examine the contribution of specific flow motions to force generation, one can use the force element method~\cite{chang1992potential} and vortex force maps~\cite{howe1995force,otomo2025vortex}, both of which rely on the introduction of an auxiliary velocity potential to derive specific aerodynamic force components.
Furthermore, scale decomposition~\cite{goto2017hierarchy,fujino2023hierarchy} isolates flow features at a specific characteristic length scale, which allows the extraction of the hierarchy of coherent vortices and the quantification of scale-dependent energy transfer.
This scale-decomposition enables extracting coherent structures under an extreme vortex-gust encounter~\cite{fukami2025extreme}.}

{The growing availability of high-fidelity numerical and experimental data has motivated the use of data-driven techniques to uncover the underlying flow physics~\cite{brunton2020machine, fukami2020assessment, taira2025machine}.
For extreme aerodynamic scenarios, where gust-induced vortex dynamics are highly unsteady and transient, nonlinear machine learning can be employed~\cite{smith2024cyclic,fukami2024data}.
Furthermore, linear data-driven modal analyses with time-varying bases have been introduced to analyze these highly transient phenomena~\cite{ashtiani2025data}.
Such techniques can provide interpretable low-order representations or modal structures, which have been challenging to achieve with traditional linear approaches due to the transient nature~\cite{linot2025extracting, taira2026extreme}.}

{
Building upon these advancements in data-driven techniques, the integration of information theory has paved the way for causality-inspired analysis for fluid flows~\cite{lozano2022information}. 
Depending on the target variable selected, information-theory-supported techniques allow for the targeted extraction of important coherent structures across diverse flow configurations. 
Arranz and Lozano-Dur{\'a}n~\cite{arranz2024informative} introduced informative and non-informative decompositions, isolating flow features responsible for wall shear stress in turbulent channel flow.
The informativeness with respect to the flow prediction has also been examined~\cite{cremades2025additive, cremades2025classically, cremades2026assessment}.
Moreover, unsteady aerodynamic flows have been examined with a causality-based data-driven technique, enabling the extraction of specific vortical motions that significantly contribute to future lift coefficients~\cite{fukami2026information, Koshikawa_Araki_Liu_Fukami_2026}.
Such a technique can be applied to industrial turbulence of flow around a vehicle with a sunroof~\cite{koshikawa2027AISIN}.}

{This study aims to extract causally important vortical structures with respect to  future target phenomena of extreme vortex gust-airfoil interaction at $Re =5000$.
This is achieved with information-theoretic convolutional causal learning~\cite{Koshikawa_Araki_Liu_Fukami_2026} that integrates a Shannon-entropy-based causality into data-driven structural extraction from a given flow snapshot. 
Applied to the turbulent wake subjected to strong disturbance, the present method isolates specific vortical structures responsible for arbitrary target phenomena in the future.
The current technique is first used for the relationship between vortical structures and the lift coefficient, capturing the contribution of large-scale structures to future aerodynamic loads.
The structural contribution to future aerodynamic coefficients measured with the present approach is examined by comparing it to force element analysis that identifies instantaneous lift-generating structures.
We further discuss the dependence of the identified informative vortical structures on the future target variable by additionally considering a scale-to-scale energy transfer.}

The present paper is organized as follows.
The formulation of informative mode decomposition and the data curation are described in section \ref{sec:method}. 
Results are discussed in section \ref{sec:res}. Conclusions are remarked in section \ref{sec:conc}.

\section{Approach}
\label{sec:method}
\begin{figure}
    \begin{center}
        \includegraphics[width=1\textwidth]{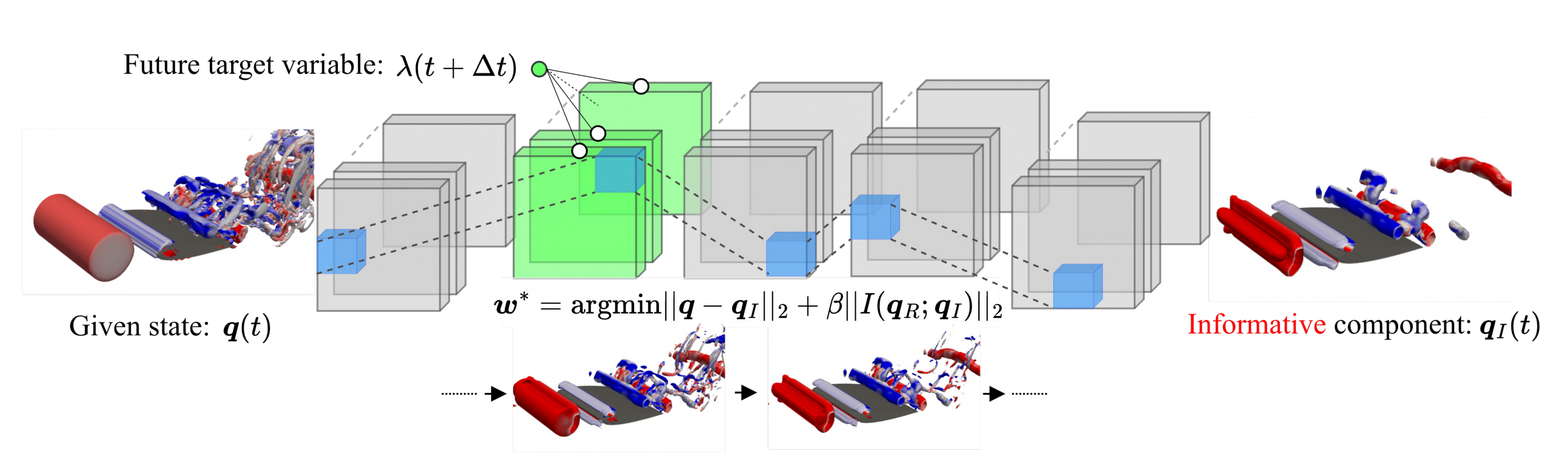}
    \end{center}
    \vspace{-2mm}
    \caption{
    An example of the given state $\bm q$ and the informative component $\bm q _I$ decomposed by the present informative mode extractor $\mathcal F$.
    }
    \vspace{-3mm}
    \label{fig1}
\end{figure}

{This study considers extracting causally important vortical structures of transient turbulent airfoil wake-strong vortex-gust interactions.
The importance of vortical structures in this study is defined based on the information-theoretic causality measure obtained from a vortical flow snapshot and future aerodynamic variables.
To build a structural extractor that accommodates the aforementioned causality-inspired concept, we consider information-theoretic machine learning based on three-dimensional convolutional neural networks, introduced in \S\ref{subsec:IMD}.
Details on the current data curation for the extreme vortex gust-airfoil interactions at $Re = 5000$ are provided in \S\ref{subsec:Data}.}

\subsection{Informative mode decomposition}
\label{subsec:IMD}
{To extract time-varying causal vortical structure from extreme aerodynamic flows, we consider information-theoretic convolutional learning for mode decomposition.
that decomposes a given state $\bm q(\bm x, t)$ based on the contribution to the future target variable $\lambda$ in the operation described as,
\begin{equation}
    \bm{q}(\bm x, t) = \bm{q}_I(\bm x, t) +\bm{q}_R(\bm x, t), 
\end{equation}
where ${\bm{q}}_{I}$ and ${\bm{q}}_{R}$ are the informative and residual components, respectively~\cite{arranz2024informative, fukami2026information,Koshikawa_Araki_Liu_Fukami_2026}.}

{This decomposition is formulated within an information-theoretic framework to extract spatial structures that possess the maximum information with respect to the future target state. 
Specifically, the informative component $\bm{q}_I$ is extracted throughout the optimization that maximizes the mutual information $I(\bm{\lambda}; \bm{q}_I) = H(\bm{\lambda}) - H(\bm{\lambda} | \bm{q}_I)$, where $H(\bm{\lambda})$ and $H(\bm{\lambda} | \bm{q}_I)$ denote the Shannon entropy and the conditional Shannon entropy, respectively~\cite{shannon1948mathematical}.
Concurrently, to isolate specific informative components from the given state, the residual component $\bm{q}_R$ is constrained to be statistically independent of $\bm{q}_I$, which enforces their mutual information to be zero, $I(\bm{q}_R; \bm{q}_I) = 0$. 
In other words, the informative components are a subset of the given state, capable of explaining the future target variable.
The comprehensive mathematical formulations are built upon informative and non-informative decomposition, introduced by Arranz, G. and Lozano-Dur{\'a}n~\cite{arranz2024informative}.}

{We consider the field of the second invariant $Q$ on the velocity gradient tensor~\cite{hunt1988eddies} as the given state to deepen our understanding of vortical motion during the extreme vortex gust-airfoil interaction.
As target variables $\bm{\lambda}$, we employ the lift coefficient $C_L \equiv F_L/(0.5\rho u_\infty ^2 c)$ and scale-dependent energy transfer $\mathcal T$, which will be detailed later.
Here, $F_L$, $u_\infty$, $c$, and $\rho$ denote the lift force, freestream velocity, chord length, and fluid density, respectively.
The current selection of the target variables is motivated to capture the important structures based on the pre-defined causality in the extreme vortex-airfoil interaction from two key physical aspects: aerodynamic lift response and scale-dependent energy transfer.}

{The present decomposition is performed with the informative mode extractor $\mathcal F$,
\begin{align}
    \bm q_I(t) = \mathcal{F}(\lambda(t+\Delta t), \bm q(t); \bm w),
    \label{eq:F}
\end{align} 
implemented with deep sigmoidal flow~\cite{huang2018neural}, as illustrated in figure~\ref{fig1}.
The optimization for the weight parameter $\bm w$ is performed with 
\begin{align}
    \bm w^* = \operatorname{argmin}_{\bm w}||\bm q - \bm q_I||_2+\beta||I(\bm q_R; \bm q_I)||_2,
    \label{eq:costfunc}
\end{align}
where the regular regression loss and mutual information loss are balanced with a constant parameter $\beta$~\cite{arranz2024informative}.
To solve this optimization problem, the Adam optimizer~\cite{kingma2014adam} is employed, and an early stopping criterion is incorporated during the training process to mitigate overfitting.
The value of $\beta$ is determined based on the L-curve analysis~\cite{hansen1993use,fukami2025compact}, facilitating the identification of the trade-off relationship between the two terms.}

\begin{figure}
    \begin{center}
        \includegraphics[width=1\textwidth]{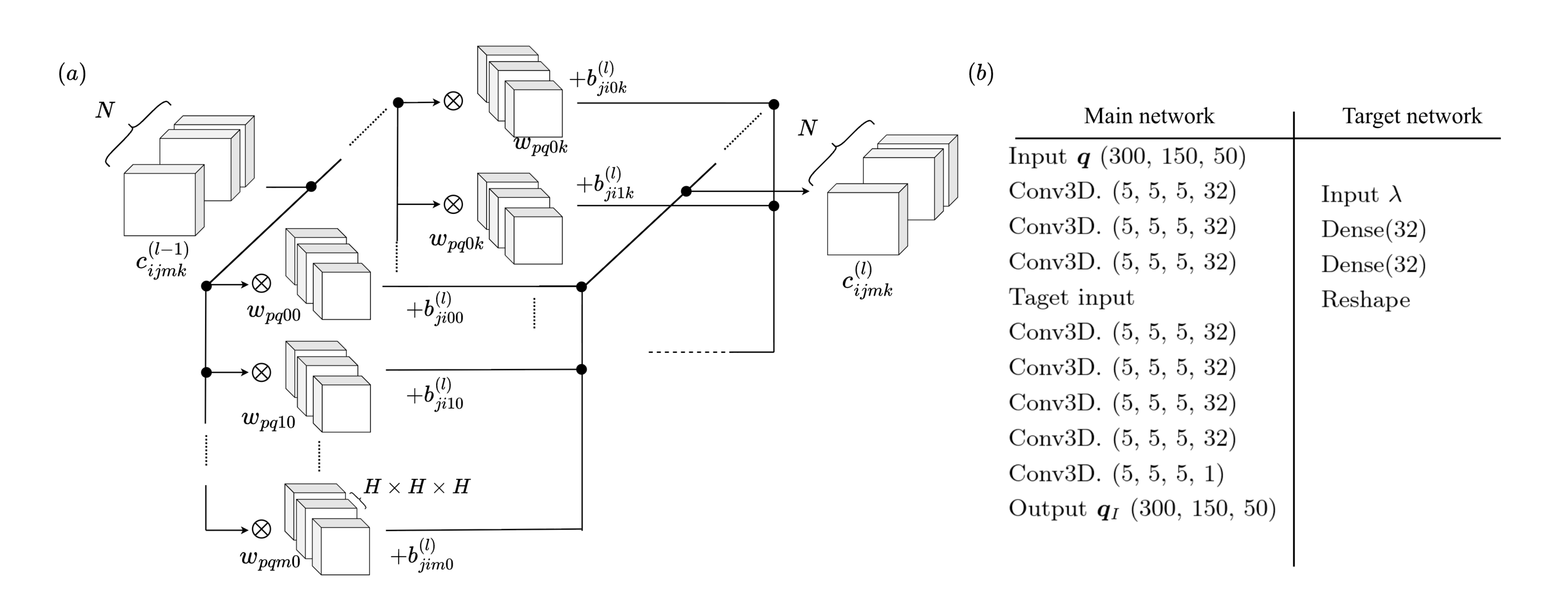}
    \end{center}
    \vspace{-2mm}
    \caption{
    ($a$)~The operation of three-dimensional convolutional layer. ($b$)~The architecture of the convolutional deep sigmoidal flow used in the present study. The convolutional layer is denoted as `Conv3D.' 
  The size of the filter $H$ and its number $N$ are shown for each layer in the form of $(H,H,H,N)$.}
    
    \vspace{-3mm}
    \label{fig_CNN}
\end{figure}

{The present study employs three-dimensional convolutional neural networks~\cite{lecun2002gradient} as an informative mode extractor, which enables the extraction of features across the entire snapshot while preserving the complex spatial coherence and arrangement of vortical structures~\cite{morimoto2021convolutional,Koshikawa_Araki_Liu_Fukami_2026}. 
The main operation of the CNN, referred to as the convolutional operation, hierarchically extracts the three-dimensional spatial features of the input data through localized filtering. 
The output of the $(l-1)$~th layer is passed to the $l$~th convolutional layer, where it is convolved with the learnable weight parameter $w_{pqrnm}^{(l)}$. 
After the addition of a bias term $b_{ijkm}^{(l)}$, an activation function $\varphi$ is applied to introduce nonlinearity. 
This series of operations for extracting flow features is expressed as
\begin{align}
    c_{ijkm}^{(l)} = \varphi \left( \sum_{n=0}^{N-1} \sum_{p=0}^{H-1} \sum_{q=0}^{H-1} \sum_{r=0}^{H-1} w_{pqrnm}^{(l)} c_{i+p-G, j+q-G, k+r-G, n}^{(l-1)} + b_{ijkm}^{(l)} \right),
    \label{eq:CNN}
\end{align}
where $G=\text{floor}(H/2)$, $H$ is the size of the cubic filter, and $N$ represents the number of input channels, as illustrated in figure~\ref{fig_CNN}($a$).
To constrain the model to provide the output $\bm q_I$ that satisfies one of the information-theoretic requirements of $H(\lambda|\bm q_I) = 0$, the weight parameters are restricted to non-negative values in conjunction with bijective activation functions~\cite{huang2018neural}.}

{We construct informative mode extractors based on the three-dimensional convolutional neural network. 
The detailed architecture of the neural network is summarized in figure~\ref{fig_CNN}($b$). 
The model takes two inputs, a given state $\bm q(t)$ and a future target variable $\lambda (t+\Delta t)$, and outputs the informative component $\bm q_I(t)$. 
Note that the present network architecture is sufficiently deep and expressive to reconstruct the entire field if the model focuses on the reconstruction loss in equation~\ref{eq:costfunc}.
In other words, the structures removed by the present model are discarded as they do not contribute to the target phenomenon, not because of the model capability.
The informative structures are finally obtained as an output of the present neural network. 
As described in equation~\ref{eq:F}, the decomposed modal structures at time $t$ are a function of two terms: the target variable $\lambda$ and time delay $\Delta t$. 
The former corresponds to the effect in a cause-and-effect pair, while the latter represents the time difference considered between them. 
We will discuss how the informative modes vary with these user-defined parameters later.}

\begin{figure}
    \begin{center}
        \includegraphics[width=0.98\textwidth]{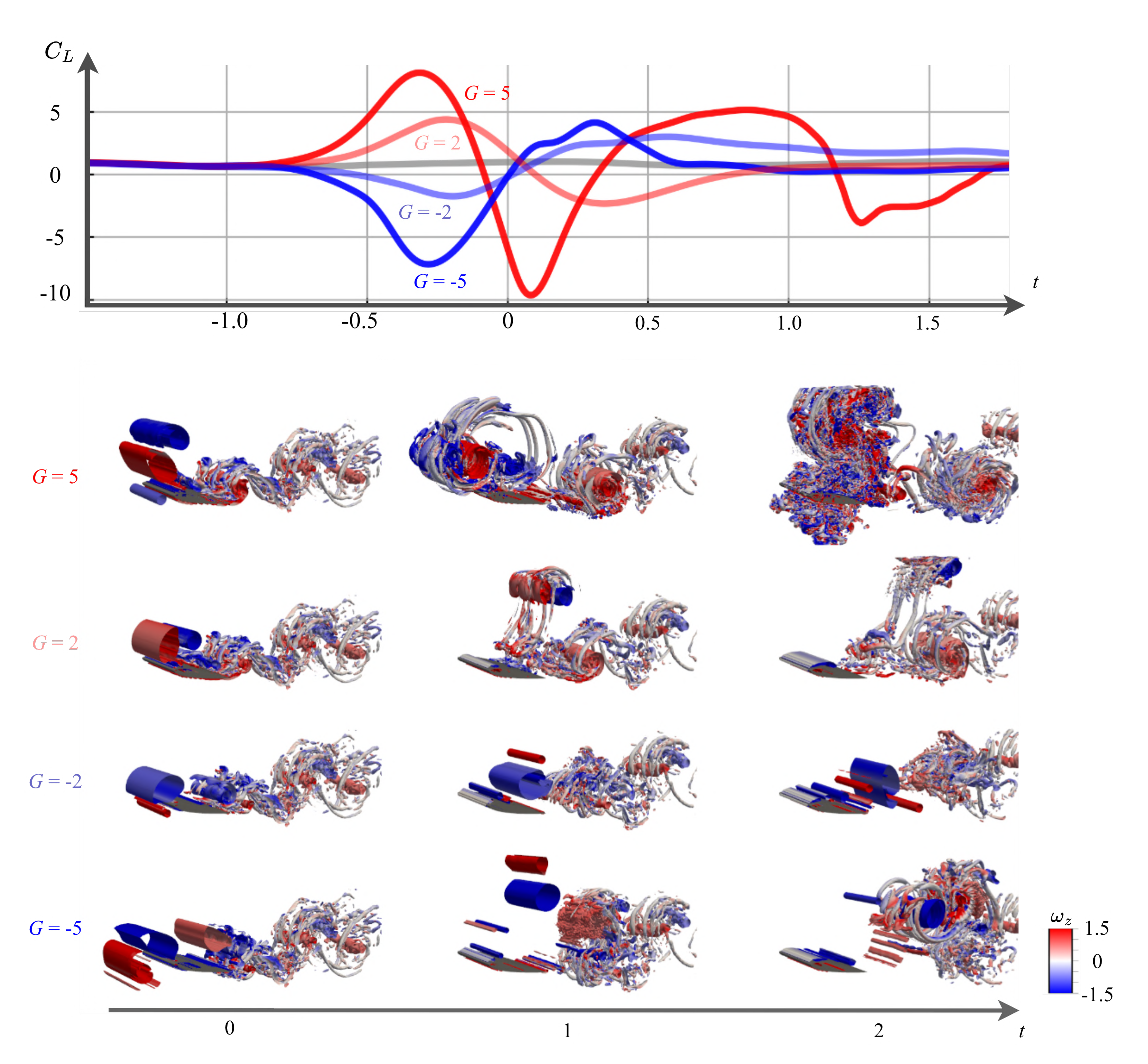}
    \end{center}
    \vspace{-2mm}
    \caption{
    Extreme vortex gust-airfoil interaction at $Re = 5000$ with ($a$) $G = 2$ and $G = 5$.
    Temporal evolution of the lift coefficient $C_L$ (gray: the undisturbed case) and the corresponding vortical structures visualized by the isocontours of $Q$-criterion ($Q_{\rm th} = 0.1$), colored by spanwise vorticity~$\omega_z$.
    }
    \vspace{-6mm}
    \label{fig:case}
\end{figure}

\subsection{Extreme vortex gust-airfoil interaction at Reynolds number of 5000}
\label{subsec:Data}

{To prepare the data set for the current analysis, we perform large-eddy simulations (LESs) under the condition of an incompressible flow around an airfoil at a chord-based Reynolds number of $Re = u_\infty c/\nu = 5000$, where $\nu$ is the kinematic viscosity. 
The angle of attack is fixed at $\alpha = 14^\circ$. 
Under the current angle of attack, the turbulent separated wake is generated, enabling the analysis of the interaction between turbulent structures and a vortex gust~\cite{fukami2025extreme}.
The simulations are carried out using the finite-volume solver Cliff, with a second-order accurate discretization scheme in both space and time~\cite{ham2004energy,ham2006accurate}. 
The computational domain extends over $(x,y,z)/c \in [-20, 25] \times [-20, 20] \times[0,1] $, with the origin located at the leading edge of the airfoil. 
For the boundary conditions, a freestream velocity is specified at the inlet and far-field boundaries, a convective boundary condition is applied at the outlet, and a no-slip condition is enforced on the airfoil surface. 
In the spanwise direction, a periodic boundary condition is applied across the domain length of $1c$ to capture the three-dimensional turbulent structures. }

{To model the extreme aerodynamic disturbance, a discrete vortex gust is introduced upstream of the airfoil at an initial location of $x_0/c = -2$.
The vortex gust is modeled as a Taylor vortex~\cite{taylor1918dissipation} whose rotational velocity $u_\theta$ is defined as 
\begin{equation}
   u_{\theta} = u_{\theta, \rm max} \frac{r}{R} \exp \Big[\frac{1}{2} \Big( 1-\frac{r^2}{R^2}\Big)\Big] ,
\end{equation}
where $r$ is the radial distance from the vortex center, $u_{\theta,\rm max}$ is the maximum tangential velocity, and $R$ is the radius at which this maximum velocity is attained.
Further details of the data curation and the numerical setup are seen in Fukami et al.~\cite{fukami2025extreme}.}

{To focus on the vortical interactions near the airfoil, for the present data-driven causality-inspired analysis, we extract the subdomain of $(x,y,z)/c \in [-1.5, 4.5] \times [-1.5, 1.5] \times [0,1]$. 
Moreover, the unstructured LES datasets within the subdomain are interpolated onto a Cartesian, uniform grid with the spatial grid resolution of $(N_x, N_y, N_z) = (300, 150, 50)$, where $N_x$, $N_y$, and $N_z$ correspond to the number of grid points along the streamwise, wall-normal, and spanwise axes, respectively, to leverage a convolutional layer inside the informative mode extractor~\cite{lecun2002gradient,Koshikawa_Araki_Liu_Fukami_2026}.
We set $t = 0$ when the center of the introduced vortex gust arrives at the location of the leading edge $x = 0$.}

\section{Results}
\label{sec:res}

{The encounter between an extreme vortex gust and an airfoil triggers swift and intense flow modifications, the nature of which varies significantly depending on the physical parameters of the incoming disturbance. 
Among these parameters, the gust ratio serves as a particularly crucial factor, as it fundamentally dictates both the relative magnitude and the rotational direction of the vortex.
To cover a variety of aerodynamic behaviors induced by vortex gust, we consider four cases with a range of gust ratios, specifically $G = \pm\{ 2,5\}$.
The time series of lift coefficient $C_L$ and the corresponding vortical flow evolutions visualized by the $Q$-criterion exhibit highly unsteady and transient dynamics dependent on the gust ratio, as shown in figure~\ref{fig:case}.
When a positive gust approaches the airfoil, the lift coefficient undergoes a sharp surge followed by a severe drop before gradually recovering toward the baseline trajectory. 
A similar, yet inverted, sequence of events is observed with negative gusts: the lift experiences a drastic initial plunge, followed by a sharp transient recovery, and converges back to the undisturbed state.
As the magnitude of $G$ becomes large, the fluctuation from the undisturbed flow increases.}

{While the time trace of lift coefficient $C_L$ shares the aforementioned symmetrical characteristics, how the vortex gust interacts with the flow around an airfoil and how vortical structures detach differ depending on the gust magnitude and sign.
When the gust with $G = 2$ is introduced to the airfoil, the vortical structures remain relatively cohesive.
Following the interaction with the airfoil, the vortical structures induced by positive gusts tend to detach and convect vertically upwards. 
Conversely, in the case of negative gusts, the structures are predominantly convected towards the trailing edge, as shown in figure~\ref{fig:case}.

Such intense flow modifications observed at higher gust ratios can be discussed via the vorticity generation due to the approaching gust near the leading edge.
The wall-normal vorticity flux, $J_n$, which quantifies the continuous injection of vorticity into the fluid~\cite{hornung1989vorticity, wu2012vorticity}, is expressed as
\begin{equation}
    J_n = -\hat{\boldsymbol{\tau}} \cdot \frac{d\boldsymbol{V}_{w}}{dt} - \frac{1}{\rho} \left( \frac{\partial p}{\partial \tau} \right)_{w},
    \label{eq:Taira}
\end{equation}
where $\bm V_w$ is the wall velocity, $p$ is the pressure, and $\bm \tau$ are the wall-normal
and tangential directions, respectively.}

{In the case of an intense positive gust, the approaching vortex core induces an extreme adverse pressure gradient along the leading edge.
This intense tangential pressure gradient dictates a massive injection of vorticity flux from the airfoil, driving immediate and extensive flow separation, following equation~\ref{eq:Taira}.
Driven by the strong rotation of the primary gust at $G = 5$, these massively separated structures subsequently re-impinge on the wing surface, triggering instabilities that generate three-dimensional turbulence with a range of scale lengths.}

{On the other hand, during the interaction with the intense negative gust at $G = -5$, the clockwise rotation of the impinging gust similarly imposes an extremely large tangential pressure gradient at the leading edge, inevitably triggering a massive injection of vorticity flux. 
In contrast to the positive gust, the induced velocity field prevents these generated secondary structures from lingering over the suction surface. Instead, the vortical structures rapidly detach and convect downstream, actively mixing with the pre-existing wake from the trailing edge.
These observations highlight that the lift generation mechanics are governed by highly transient and swift vortical interaction, where traditional correlation-based modal analyses may fail to isolate the driving mechanisms.
These observations based on the vorticity flux are consistent with what we could find from the pressure coefficient $C_p$ on the wing surface~\cite{fukami2025extreme}.}

{While the na\"ive analyses based on flow contours and time traces of aerodynamic coefficients above can still provide physical insights into transient aerodynamics, of interest here is whether they can be further evidenced and a new mechanism would be discovered from the information theoretic viewpoint.
Let us first apply the present time-varying structural extraction to obtain the informative components contributing to the future lift coefficient across the four gust ratios considered herein.
The present approach extracts specific vortical structures from gust-ratio-dependent interaction mechanisms described above.
We then examine these extreme interactions through the lens of the inter-scale energy transfer. 
By decomposing the flow field based on the energy transferred across different spatial scales, we isolate the important vortical structures responsible for driving inter-scale interaction. 
Finally, we examine the dependence of the extracted informative modes on the user-defined target variables, which enables the discussion about the relationship between the lift-based mechanism and scale-dependent energy transfer.}

\subsection{Informative structures with respect to the future lift coefficient}

{Let us consider informative mode decomposition with respect to future lift response.
Viewing a cause-and-effect relationship between the current vortical structure and the future lift response, the present model aims to extract transient lift-related structures as informative modes.
While the present decomposition extracts informative vortical structures based on the mutual information with the future lift coefficient, one can also physically identify the vortical structures responsible for lift generation based on the force element analysis~\cite{chang1992potential}. 
An auxiliary potential $\phi_y$ is defined for a boundary condition of $-\boldsymbol{n} \cdot \nabla \phi_y = \boldsymbol{n} \cdot \boldsymbol{e}_y$ on the wing surface. Here, $\boldsymbol{e}_y$ is the unit vector in the y direction. 
Integrating the inner product of the Navier--Stokes equations with the gradient of the auxiliary potential over the fluid domain, the lift force can be expressed as
\begin{equation}
    F_L = \int_V \boldsymbol{\omega} \times \boldsymbol{u} \cdot \nabla \phi_y dV + \frac{1}{Re} \int_S \boldsymbol{\omega} \times \boldsymbol{n} \cdot (\nabla \phi_y + \boldsymbol{e}_y) dS ,   
\end{equation}
where the first and second terms represent volume and surface lift elements, respectively. 
In the current separated wake flow at $Re = 5000$, the volume lift element becomes dominant compared to the surface element~\cite{ribeiro2023laminar, odaka2026extremeb, odaka2026vorticala}.
As such, the volume lift element is hereafter referred to as lift element $L_e$, which serves as a metric to evaluate the data-driven, causality-inspired informative modes extracted by our machine learning framework.
While the lift element analysis provides the spatial location of flow structures that generate the lift force at a given instant, the informative mode decomposition extracts the informative structures with respect to the future lift response after a time delay $\Delta t$, such that
\begin{equation}
    \bm q_{I,C_L}(t) =  \mathcal{F}(\bm q(t), C_L(t+\Delta t)),
\end{equation}
where $\bm q_{I,C_L}$ is the decomposed informative components with respect to the lift response.
We will discuss the dependence of informative modes on $\Delta t$ later.}

\begin{figure}
    \begin{center}
        \includegraphics[width=1\textwidth]{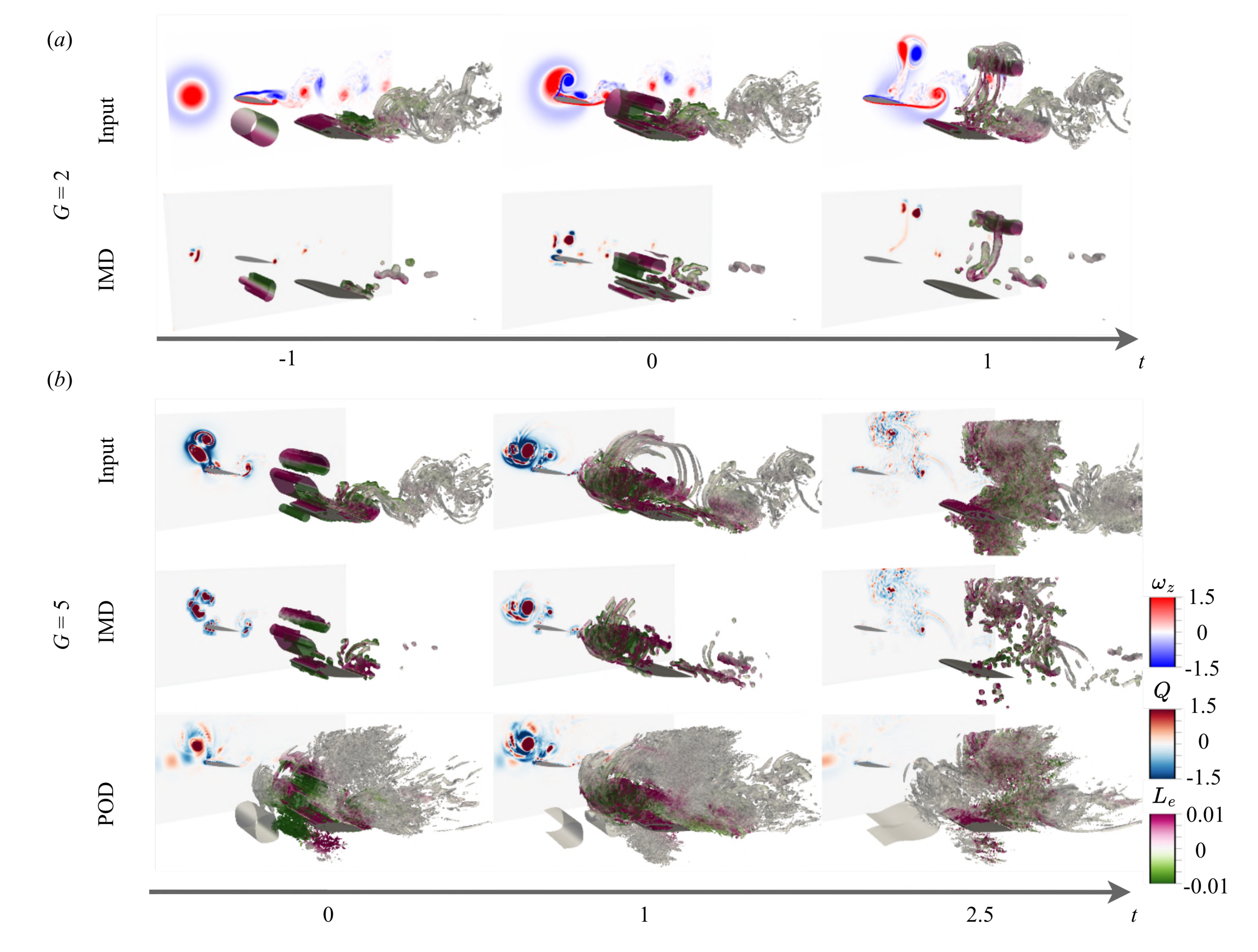}
    \end{center}
    \vspace{-2mm}
    \caption{
    Informative mode decomposition for extreme positive vortex gust encounter at $Re = 5000$ with ($a$) $G = 2$ and $G = 5$. The iso-surfaces of $Q$-criterion ($Q_{\rm th} = 0.1$) colored by lift element $L_e$ are visualized in the foreground. The background is spanwise vorticity $\omega_z$ and spanwise-averaged decomposed $Q$-criterion fields. The bottom row is the flow field reconstructed using 12 proper orthogonal decomposition modes applied to the $Q$-criterion field.
    }
    \vspace{-3mm}
    \label{fig:IMD_pos}
\end{figure}

{Let us begin with informative mode decomposition for the positive gust-airfoil interaction with  $G = 2$.
Here, the time window $\Delta t$ is set to $0.05$.
The current model extracts specific vortical structures responsible for the lift coefficient, as shown in figure~\ref{fig:IMD_pos}($a$).
At $t = -1$, the model extracts two distinct flow structures as informative: the large-scale vortex cores of the separated wake vortical structures, and the airfoil-side component of the approaching vortex gust, while fine-scale vortical structures are filtered out from the extracted modes.
Notably, this exclusion is not achieved by explicitly providing spatial scale information to the model. 
Rather, it is determined based on their lack of contribution to the future lift coefficient~\cite{Koshikawa_Araki_Liu_Fukami_2026}.
When the vortex core impacts the leading edge at $t = 0$, the model extracts the rolled-up gust structures near the leading edge and the vortical features near the trailing edge. 
Furthermore, this extraction exhibits qualitative agreement with the spatial location of lift-generating structures obtained from the force elemental method.
In other words, the vortical structures decomposed with the current data-driven method are informative with respect to the lift generation mechanism.}

{The informative modes are not limited solely to structures associated with rotating vortex cores. At $t = 1$, the model identifies the structures generated as the vortex gust lifts off upward from the suction side as highly informative.
This indicates that these isolated structures are not merely rotational signatures, but rather capture a distinct physical mechanism linking the flow dynamics to lift generation, extending beyond the conventional framework of circulation.}

\begin{figure}
    \begin{center}
        \includegraphics[width=1\textwidth]{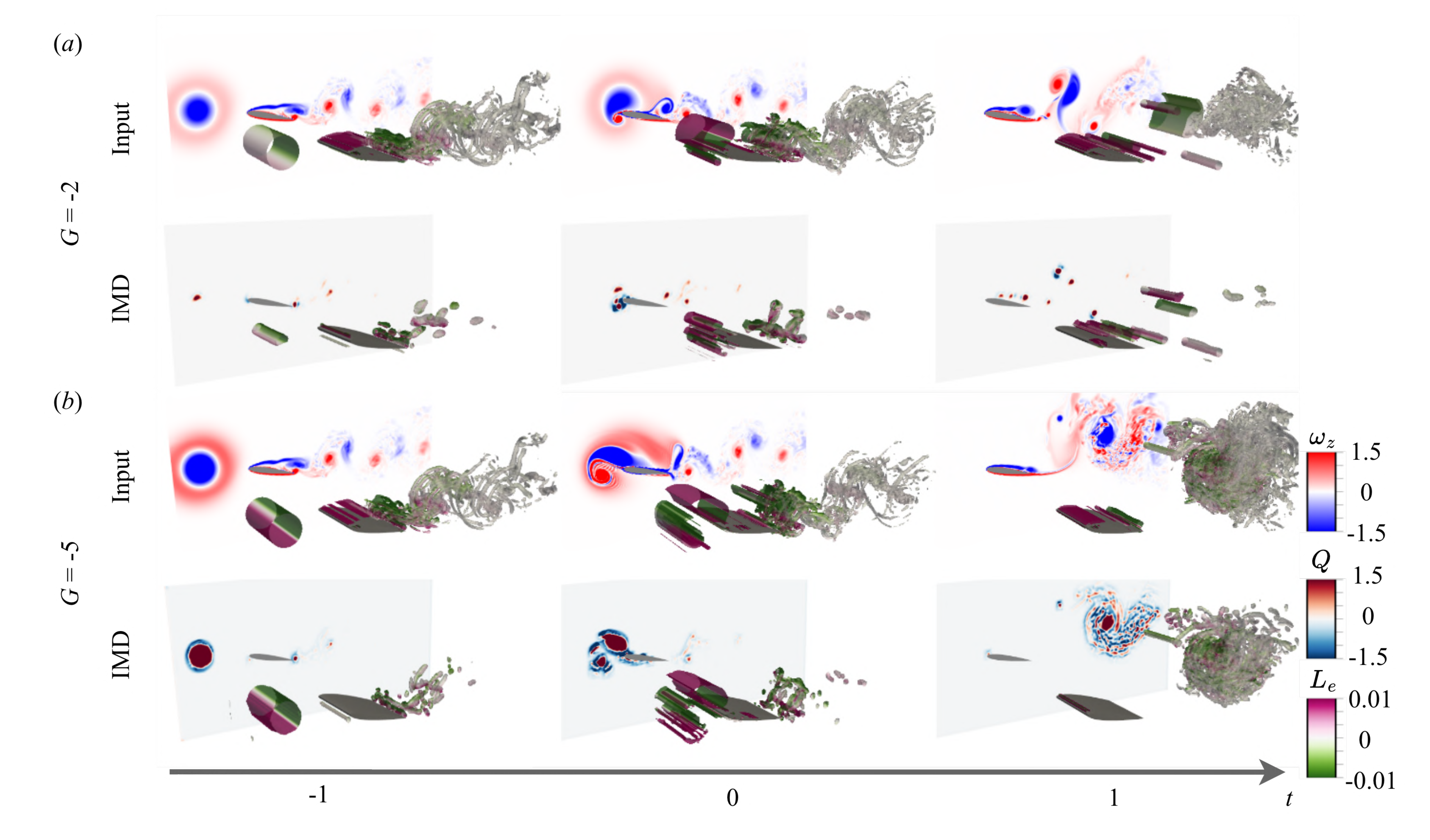}
    \end{center}
    \vspace{-2mm}
    \caption{
    Informative mode decomposition for extreme negative vortex gust encounter at $Re = 5000$ with ($a$) $ G = -2$ and ($b$) $G = -5$. The iso-surfaces of $Q$-criterion ($Q_{\rm th} = 0.1$) colored by lift element $L_e$ are visualized in the foreground. The backgrounds are spanwise vorticity $\omega_z$ and spanwise-averaged decomposed $Q$-criterion fields.
    }
    \vspace{-3mm}
    \label{fig:IMD_neg}
\end{figure}

{To examine the cause-and-effect relationship under a more intense disturbance, we next perform the present decomposition on the case of $G = 5$, which induces a swifter excitation of lift response, as shown in figure~\ref{fig:IMD_pos}($b$).
During the period spanning from the first to the second impingement, the gust structures associated with both impacts are dominantly extracted as informative modes, which exhibit a strong spatial correspondence with the dominant lift elements.
In contrast to the earlier stages of the interaction, where large-scale coherent structures are primarily extracted, at $t = 2.5$, the model selectively isolates specific features from a cluster of mid-scale turbulent structures generated over the suction side of the airfoil. 
These extracted informative modes are not solely confined to vortex cores with intense rotation. 
Instead, the model additionally identifies the elongated structures continuously connected to the rolling-up vortical structures, which exhibit qualitative analogy to the decomposed structures observed in the case of $G =2$, presented in figure~\ref{fig:IMD_pos}($a$).
Such a structural similarity indicates that, although the flow evolution and the severity of the breakdown vary significantly depending on the gust ratio, the underlying physical mechanisms driving the instantaneous aerodynamic force remain universal.}

{To emphasize that the present approach extracts specific flow features based on their causal dominance rather than mere statistical variance, we compare our results with proper orthogonal decomposition~\cite[POD;][]{lumley1967structure}.
POD extracts spatial modes that capture the statistical variance based on the energy perspective.
We use the leading 12 POD modes to reconstruct the dominant features, as shown in figure~\ref{fig:IMD_pos}($b$). 
Note that this selected number of modes is sufficiently large to capture the coherent structures of the approaching vortex gust. 
Following the moment of gust impingement at $ t= 1$, the POD reconstruction prominently exhibits large-scale wake structures that are regarded as less informative by the present informative modes.
Moreover, it introduces non-physical artifacts arising from the intense gust disturbance lingering in the statistical mean flow. 
This contrast suggests that the present causality-inspired mode decomposition successfully filters out energetically dominant yet causally irrelevant features, isolating structures based on their contribution to the future lift rather than their kinematic energy magnitude.}

{We next consider the interaction with a negative vortex gust, as shown in figure~\ref{fig:IMD_neg}. 
Similar to the positive cases, the model extracts lift-associated vortical structures, while adapting to the magnitude of the disturbance. 
At $t = -1$, a localized portion of the gust is identified as informative for $G = -2$, whereas the majority of the gust structure is extracted for $G = -5$. 
This difference agrees with the lift element distribution, which indicates a relatively minor lift contribution from the outer region of the gust at $G = -2$. 
Subsequently, at $t = 0$, the model isolates the informative vortical structures highlighted by the lift elements in the vicinity of the airfoil, which is similar to what is observed when we introduce the positive gust.
Furthermore, as the interaction progresses, the model tracks the large-scale vortical structures detaching from the airfoil. 
These consistent extractions across different gust signs and magnitudes suggest the capability of the present causality-inspired approach to robustly isolate the informative flow features from a diverse range of extreme aerodynamic interactions.}

\begin{figure}
    \begin{center}
        \includegraphics[width=1\textwidth]{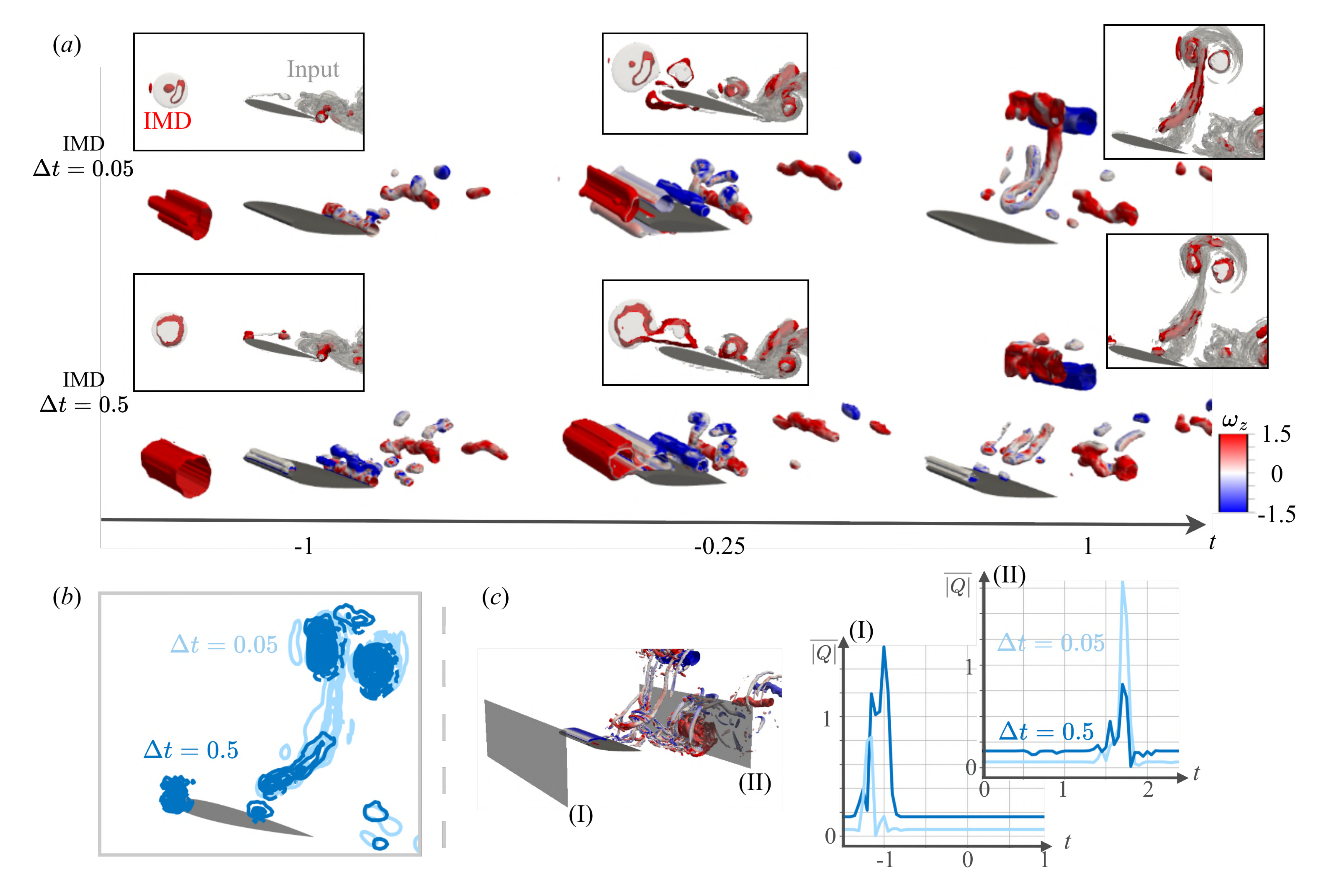}
    \end{center}
    \vspace{-2mm}
    \caption{
    The dependence of informative modes on the value of the time window $\Delta t$. ($a$)~Informative components are visualized with an iso-surface of $Q$-criterion ($Q_{\rm th}$ = 0.1). ($b$)~The zoomed-in view of the extracted mode at $t = 1$. Two informative modes are superposed on the spanwise-averaged two-dimensional contour field.
    ($c$)~The time series of the magnitude of $Q$-criterion field on two $y-z$ planes~(I: $x = -1$, I\hspace{-1.2pt}I: $x=2$).
    }
    \vspace{-3mm}
    \label{fig:Delta_t}
\end{figure}

\begin{figure}
    \begin{center}
        \includegraphics[width=1\textwidth]{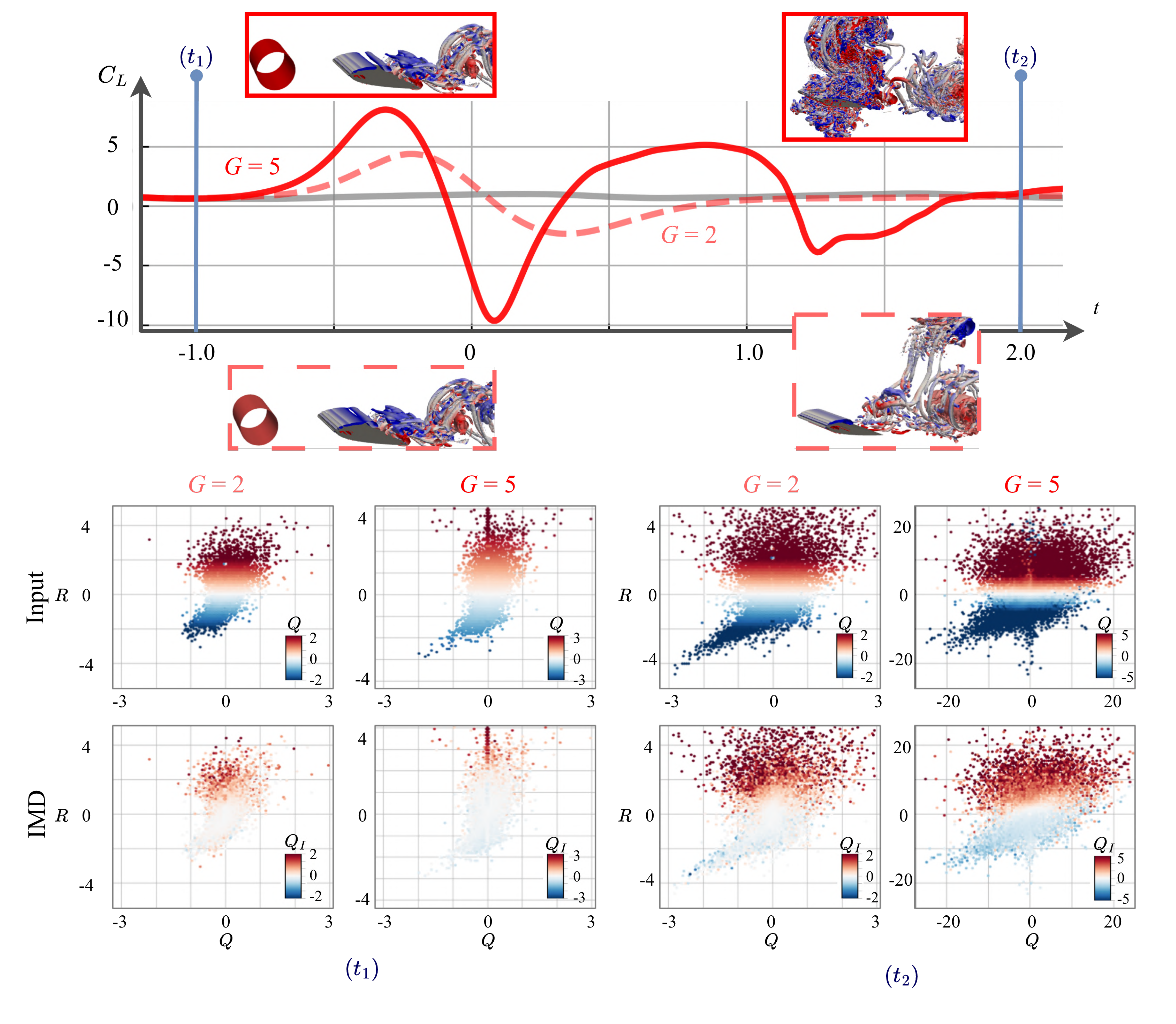}
    \end{center}
    \vspace{-2mm}
    \caption{
    Time-dependent $Q$-$R$ distributions of extreme vortex-gust airfoil interaction wake for gust ratios of $G = 2$ and $5$, colored by input and informative $Q$-criterion.
    }
    \vspace{-3mm}
    \label{fig:Q_R}
\end{figure}

{Let us examine the dependence of informative modes on the value of the time window $\Delta t$. While lift element analysis identifies flow structures at time $t$ that generate lift at that exact same time $t$, the present causality-based model is capable of extracting the specific structures at time $t$ that will causally contribute to the lift response at a future time $t + \Delta t$~\cite{Koshikawa_Araki_Liu_Fukami_2026, koshikawa2027AISIN}.
Thus, examining this temporal dependence provides an opportunity to examine how the present data-driven approach complements the force element analysis.
To demonstrate how this combined physical insight enhances our understanding of transient aerodynamics, we perform the present decomposition for the case of $G = 2$ with two distinct values of $\Delta t$, as shown in figure~\ref{fig:Delta_t}.
For a smaller $\Delta t$, a localized portion of the approaching gust from upstream is identified as informative. 
On the contrary, by accounting for a larger time delay, the entire gust structure becomes evaluated as an informative mode. 
The current time-delay effect is consistent with previous studies applying causality-based decomposition to two-dimensional extreme vortex interactions at $Re = 100$~\cite{fukami2026information, Koshikawa_Araki_Liu_Fukami_2026}.}

{In addition to the primary gust body, the choice of $\Delta t$ significantly affects the extraction of secondary structures generated by the interaction. 
For instance, at $t = 1$, the elongated structures produced by the gust detaching from the suction side are evaluated as less informative for a larger time delay contribution.
As highlighted in figure~\ref{fig:Delta_t}($b$), while the elongated shear extends continuously to the departing gust for $\Delta t = 0.05$, it is truncated to nearly half its length, remaining closer to the airfoil, for $\Delta t = 0.5$. 
The location of these truncated structures coincides with the region where interactions between structures with different length scales occur due to the gust departure, which will be discussed in detail later. 
This selective removal implies that the lift contribution from the scale-length-boarding interactions induced by the extensive modification by vortex gust impingement is highly instantaneous. }

{To quantitatively discuss the time-delay effect, the time series of the spatially averaged magnitude of the $Q$-criterion, $\overline{|Q|}(t,x) = \overline{|Q(t,x,y,z)|}^{y,z}$, is assessed in figure~\ref{fig:Delta_t}($c$). We consider two representative $y-z$ cross-sections: an upstream plane at $x = -1$ (Plane I) to evaluate the pre-impingement gust, and a downstream plane at $x = 2$ (Plane II) to capture the flow structures modified by the gust and interaction with the existing turbulent wake from the trailing edge.
At the upstream location, the informative component corresponding to the larger time delay exhibits a significantly stronger peak, indicating that the model with the focus on the larger $\Delta t$ assesses the approaching gust structures as informative compared to the model with a shorter $\Delta t$.
Conversely, at the downstream location, the trend is reversed. 
The flow structures extensively modified by the gust-airfoil interaction show a peak for the shorter $\Delta t = 0.05$.
These contrasting spatial responses capture that the present causality-based framework shifts its structural emphasis depending on the user-defined time delay of interest. 
Such a temporal selectivity enables filtering the complex vortical motion with a contribution to the lift force based on the time-delay, thereby providing an integrated spatio-temporal understanding of the extreme aerodynamic interaction.}

{To examine the topological nature of the decomposed structures across the different stages of the extreme interaction, the decomposed fields are evaluated on the $Q-R$ plane, as shown in figure~\ref{fig:Q_R}.
The second invariant $Q$ represents the local balance between fluid rotation ($Q > 0$) and strain ($Q<0$), whereas the third invariant $R$ denotes the mechanisms of vortex stretching ($R > 0$) and compression ($R < 0$)~\cite{davidson2015turbulence}.
During the approaching phase prior to impingement, the input flow topologies for both $G = 2$ and $G = 5$ exhibit the characteristic teardrop signature typically associated with coherent vortical structures~\cite{ooi1999study,fukami2024data_pi}.
At this early stage, the informative mode extraction predominantly isolates rotation-dominated structures, which is consistent with our previous study identifying informative vortical structures in a separated turbulent wake at $Re = 20,000$~\cite{Koshikawa_Araki_Liu_Fukami_2026}.
This selectivity is particularly pronounced in the severe $G = 5$ encounter, where the extracted causal structures are highly concentrated within the region with strong $Q$ and weak $R$, which corresponds to the approaching intense vortex gust. 
Conversely, as the extensively modified vortical structures depart the airfoil, the topological landscape shows a distinct transformation, reflecting the generation of fine-scale turbulence induced by the impingement. 
While the model continues to selectively capture the rotation-dominated vortex cores ($Q > 0$), it simultaneously extracts specific regions subjected to extensive strain ($Q < 0$) as informative, particularly in the case of intense gust.
This concurrent extraction suggests that the causal importance for the future lift response is no longer determined solely by isolated rotational bodies. 
Instead, the future lift response is causally driven by interacting structures, where the surviving rotational cores and the surrounding extensive shear layers collectively dictate the transient lift force.}

\subsection{Informative structures with respect to the energy transfer}
{Shifting our focus from the lift response, let us now examine the extreme vortex gust-airfoil interaction from the point of the energy transfer across structures of varying length scales.
While extreme vortex gust introduces the swift, large excitation of aerodynamic forces, they simultaneously trigger the generation of vortical structures across a range of length scales.}

{To isolate flow structures based on their characteristic length scale, the scale-decomposition analysis~\cite{goto2017hierarchy,fujino2023hierarchy} is performed.
A spatial band-pass filter based on a three-dimensional Gaussian kernel $f$ is applied to the velocity field in spatial space. 
The filtered velocity field $\boldsymbol{u}^{[\sigma_1, \sigma_2]}$ is obtained as
\begin{align}
    \boldsymbol{u}^{[\sigma_1, \sigma_2]}(x, y, z, t) &= \int_{\mathcal D} \boldsymbol{u}(x', y', z', t) \Big[ f(x', y', z'; x, y, z, \sigma_1) \nonumber \\
    &\quad - f(x', y', z'; x, y, z, \sigma_2) \Big] dx' dy' dz'
\end{align}
~where
\begin{align}
    f(x', y', z'; x, y, z, \sigma) = \frac{1}{(2\pi \sigma^2)^{3/2}}{\rm exp} \bigg[-\frac{(x-x')^2+(y-y')^2+(z-z')^2}{2\sigma^2} \bigg],
\end{align}
and $\mathcal D$ is the integration area.
The maximum cut-off length scale $\sigma_{\rm max}$ is set to $\sigma_{\rm max}=c/(2\pi St)$, which corresponds to the diameter of the primary roller vortices, where the shedding Strouhal number $St$ is determined from the lift coefficient of the undisturbed, quasi-periodic baseline flow prior to the gust encounter.}

\begin{figure}
    \begin{center}
        \includegraphics[width=1\textwidth]{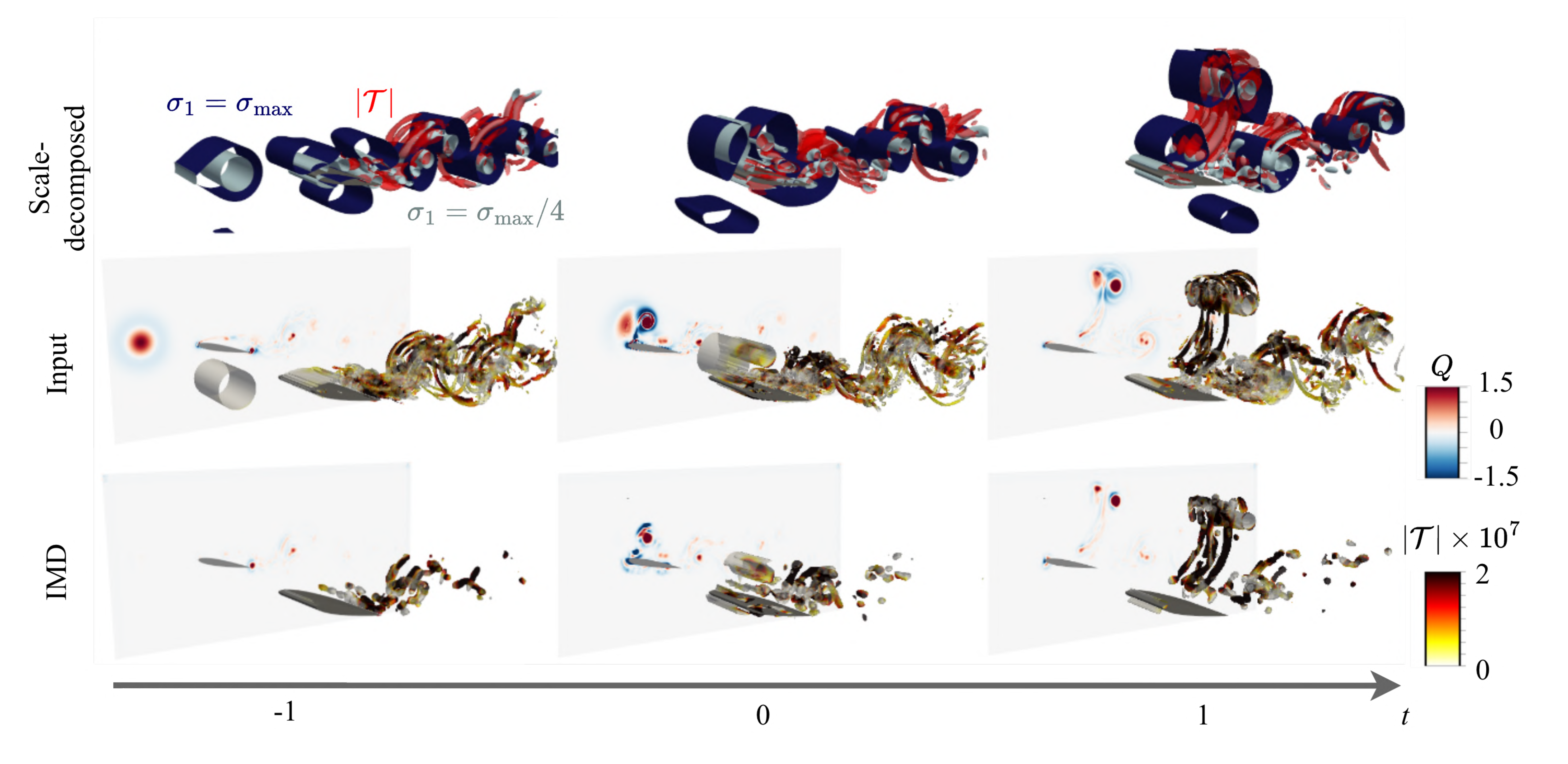}
    \end{center}
    \vspace{-2mm}
    \caption{
    Scale-decomposition and informative mode decomposition with respect to the scale-dependent energy transfer applied to extreme vortex gust encounter at $Re = 5000$.
    Shown above are the iso-surfaces of scale-decomposed $Q$ and scale-dependent energy transfer $|\mathcal T|$, while input and decomposed $Q$-criterion ($Q_{\rm th} = 0.1$) colored by $|\mathcal T|$ are visualized in the foreground. 
    Shown in the background are spanwise-averaged decomposed $Q$-criterion fields.
    }
    \vspace{-3mm}
    \label{fig:IMD_tau}
\end{figure}

{By extracting specific scale bands, the $Q$-criterion fields for a length scale range of $[\sigma_1, 2\sigma_1]$ with representative values of $\sigma_1 \in \{\sigma_{\rm max}, \sigma_{\rm max}/4\}$ are computed from the scale-decomposed velocity fields, as shown in figure~\ref{fig:IMD_tau}.
The energy transfer from a provider structures with scale $\sigma_p$ to a receiver counterpart, with scale $\sigma_r$ is then computed as
\begin{equation}
     {\mathcal T}(\sigma_p \rightarrow \sigma_r) = \omega_i^{(\sigma_r)} S_{ij}^{(\sigma_p)} \omega_j^{(\sigma_r)} \sigma_r^2,
\end{equation}
where $\omega ^{(\sigma)}$ and $S^{(\sigma)}$ represent vorticity and strain-rate tensor with the length scales of $\sigma$ evaluated with the scale-decomposition analysis~\cite{goto2017hierarchy}.}

{While the current operation provides a spatial location of instantaneous energy transfer mechanisms, the specific structures causally responsible for driving the energy transfer over a transient period remain hidden in the high-dimensional flow motion.
To isolate these underlying mechanics in a data-driven manner, we consider the time series of scale-dependent energy transfer integrated over the domain as a target variable for the informative mode decomposition, such that
\begin{equation}
    \bm q_{I,\mathcal T}(t) =  \mathcal{F}(\bm q(t), {\mathcal T}(t+\Delta t)),
\end{equation}
where $\bm q_{I,\mathcal T}$ is the decomposed informative components with respect to the transferred energy across the different length scales.}

{Let us perform the present decomposition based on the energy transfer for the vortex gust-airfoil interaction with a gust ratio of $G = 2$. 
The present model identifies specific vortical structures responsible for the transferred energy. 
At $t = -1$, the approaching main gust body is absent from the extracted informative modes, which suggests that the model captures the lack of cross-scale interactions within the gust prior to the impingement. 
Meanwhile, mid-scale features, including rib or blade structures, are selectively extracted in the wake region, highlighting the localized areas where dynamic inter-scale energy transfer is occurring.
Subsequently, at $t = 0$, when the gust impinges on the leading edge, only the left portion of the gust core is prominently identified. 
This suggests that the intense vorticity flux from the leading edge severely deforms the wall-facing side of the counter-clockwise gust, thereby triggering multiscale interactions and energy transfer.
Moreover, as the gust detaches from the suction side at $t = 1$, the decomposition highlights the elongated shear layers and the vortical cores rolling them up. 
Thus, the extraction implies that the model distinguishes the structures based on this dynamic interaction, identifying the vortical motion where the dominant contribution of the energy transfer originates.}

{The present causality-inspired framework allows the target of informativeness to be tailored to the phenomenon of interest.
Comparing the modes extracted for different target phenomena is equivalent to assessing how the informativeness of flow components shifts depending on the chosen effect within the cause-and-effect relationship.
In the present case, we have found that the two informative modes share common features at certain transient stages, such as the elongated structures on the suction side.
To quantitatively evaluate the structural similarity between the informative structures driving the transient lift dynamics and those driving the energy transfer, we compute the temporal evolution of the spatial cosine similarity, computed as
\begin{equation}
\cos\theta_I(t) = \frac{\langle \bm q_{I, C_L}(t), \bm q_{I, \mathcal{T}}(t) \rangle}{\| \bm q_{I, C_L}(t) \| \| \bm q_{I, \mathcal{T}}(t) \|} ,
\end{equation}
where ${\langle \cdot,\cdot \rangle}$ denotes the inner product.}

\begin{figure}
    \begin{center}
        \includegraphics[width=1\textwidth]{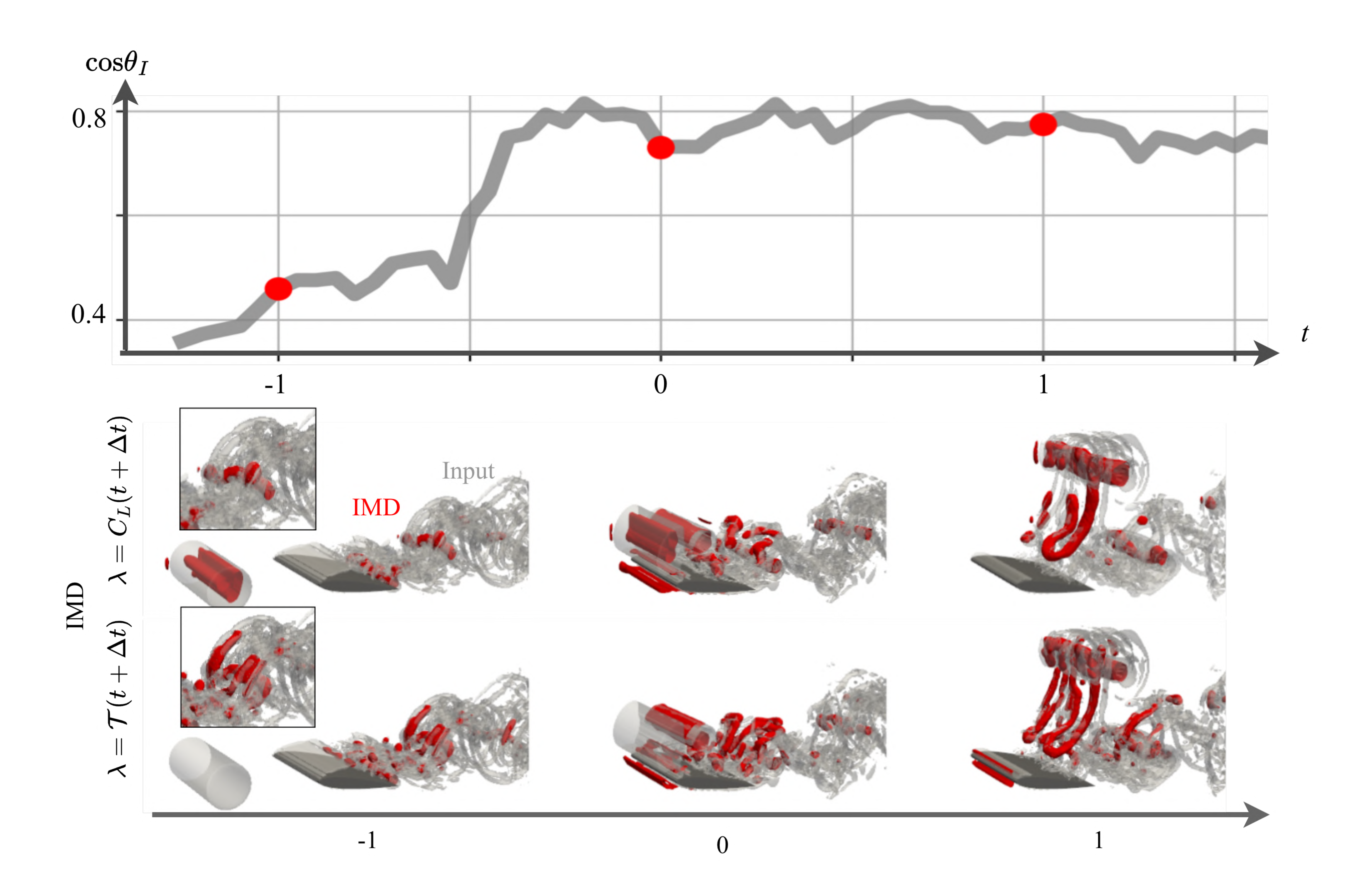}
    \end{center}
    \vspace{-2mm}
    \caption{
    Comparison of informative modes extracted with respect to distinct target variables. Shown above is the time series of the spatial cosine similarity $\cos\theta_I$. 
    Visualized below are the iso-surfaces of extracted informative components superposed onto the given $Q$-criterion field ($Q_{\rm th} = 0.1$).
    }
    \vspace{-3mm}
    \label{fig:Compare}
\end{figure}

{
We then examine how the informative modes vary with respect to the targeted physical phenomena over time. 
To highlight which structures in the given state are assessed as informative, we superpose the extracted informative components onto the given $Q$-criterion state in figure~\ref{fig:Compare}.
Note that each modal structure is the same as what is shown in figures~\ref{fig:IMD_pos} and~\ref{fig:IMD_tau}.
Prior to the gust impingement at $t = -1$, the informative structures associated with the future lift coefficient prominently contain the primary coherent core of the approaching gust and large-scale vortical cores in the turbulent wake. 
In contrast, the decomposition targeted at energy transfer filters out the undisturbed primary gust, selectively isolating mid-scale features such as rib structures in the wake. 
This distinct physical emphasis is quantitatively reflected by a relatively low spatial cosine similarity of $\cos\theta_I \approx 0.4$.}

{As the interaction progresses to $t = 0$, a spatial distinction within the gust core emerges. 
While the lift-targeted decomposition captures a broader region encompassing both sides of the impinging gust core, the energy-targeted extraction isolates only the heavily sheared proximal portion where the active inter-scale interaction occurs, which has been discussed above. 
As impingement progresses, the spatial similarity undergoes a significant increase, approaching $\cos\theta_I \approx 0.8$ during and after the impingement. 
This ascending trend suggests that the greatly deformed structures due to the impingement simultaneously drive the transient lift mechanism. 
Such an integrated physical insight suggests the applicability of the present approach to evaluate distinctive aerodynamic mechanisms based on the vortical interaction within a common field representation.}

\section{Concluding remarks}
\label{sec:conc}
{This study considered a data-driven, causality-based mode extraction on the unsteady and transient vortex dynamics of extreme gust-airfoil interactions at $Re = 5000$.
By framing the pair of cause-and-effect relationships, the present convolutional causal learning extracts specific components from the given state based on the contribution to the future target variable.
While previous studies performing the present decomposition on unsteady aerodynamic flows focus on the vortical structures related to the future lift coefficient~\cite{fukami2026information, Koshikawa_Araki_Liu_Fukami_2026}, this study examined the dependence of informative modes on the physical target phenomenon.
When targeted at the future lift coefficient, the model predominantly identified the large-scale vortical cores and approaching vortex gust as informative components.
The data-driven extraction showed qualitative spatial agreement with instantaneous force element analysis while uniquely providing the time delay between cause and effect, indicating that the present information-theoretic framework captures underlying physics from the snapshot data.
Beyond aerodynamic force response, the present framework was further applied to identify the important vortical structures of scale-dependent energy transfer. 
The decomposition selectively isolated localized regions undergoing active scale-length-broadening interactions, including the structures extensively modified by vortex gust.}

{We focused on the vortical flow around an airfoil to extract important features from turbulence with spatiotemporal complexity. 
While the force element method requires a theoretical potential flow and explicit governing equations, the present approach extracts lift-related structures solely from observable data. 
Such data-driven characteristics enable direct application to experimental flows, where resolved potential flow is not available~\cite{fukami2025observable}.
By relying on statistical mutual information, the framework can be applied to arbitrary complex dynamical systems without governing equations.
The present causality-inspired framework can extract cause-and-effect relationships from the dynamics of interest, including highly transient aerodynamic flows.}

{While the current methodology extracts informative features based on mutual information along with a neural network, the validation of physical causality could be achieved through intervention-based causal inference~\cite{pearl2009causal}.
By artificially introducing localized perturbations into the identified vortical structures and observing the subsequent temporal evolution of the dynamics, one can quantitatively evaluate the sensitivity of the future lift to these informative modes. 
Such an approach with interventional causality may provide a physical validation of the data-driven causal extractions.
Furthermore, identifying these sensitive and important structures with respect to aerodynamic forces paves the way for flow control. 
In particular, the spatial maps of the informative modes may guide the optimal placement and actuation timing, which have been performed with resolvent analysis~\cite{mckeon2010critical,ribeiro2024triglobal} and phase-amplitude analysis~\cite{fukami2024data}.
Additionally, incorporating the extracted informativeness as a localized reward function in reinforcement-learning-based active flow control frameworks presents a promising avenue for future work.}

\section*{Acknowledgements}
{
K.F. acknowledges support from the JSPS KAKENHI Grant No.~JP25K23418, the JST PRESTO Grant No.~JPMJPR25KA, and the MEXT Coordination Funds for Promoting Aerospace Utilization Grant No.~JPJ000959. 
R.A. acknowledges support from the JSPS KAKENHI Grant No.~JP24K22942 and the JST PRESTO Grant No.~JPMJPR25K1.
Q.L. acknowledges support from the US AFOSR Grant No.~FA9550-24-1-0069.
}

\section*{Declaration of interests}

{The authors report no conflict of interest.}


\bibliographystyle{unsrt}  
\bibliography{refs}

@article{arranz2024informative,
  title={Informative and non-informative decomposition of turbulent flow fields},
  author={Arranz, G. and Lozano-Dur{\'a}n, A.},
  journal={J. Fluid Mech.},
  volume={1000},
  pages={A95},
  year={2024},
  publisher={Cambridge University Press}
}

@article{chang1992potential,
  title={Potential flow and forces for incompressible viscous flow},
  author={Chang, C.-C.},
  journal={Proc. Roy. Soc. A},
  volume={437},
  number={1901},
  pages={517--525},
  year={1992},
  publisher={The Royal Society London}
}

@article{otomo2025vortex,
  title={Vortex force map method to estimate unsteady forces from snapshot flowfield measurements},
  author={{\=O}tomo, S. and Gehlert, P. and Babinsky, H. and Li, J.},
  journal={Exp. Fluids},
  volume={66},
  number={3},
  pages={1--13},
  year={2025},
  publisher={Springer}
}

@article{howe1995force,
  title={On the force and moment on a body in an incompressible fluid, with application to rigid bodies and bubbles at high and low {R}eynolds numbers},
  author={Howe, M.},
  journal={Q. J. Mech. Appl. Math.},
  volume={48},
  number={3},
  pages={401--426},
  year={1995},
  publisher={Oxford {U}niversity {P}ress}
}

@inproceedings{huang2018neural,
  title={Neural autoregressive flows},
  author={Huang, C.-W. and Krueger, D. and Lacoste, A. and Courville, A.},
  booktitle={International conference on machine learning},
  pages={2078--2087},
  year={2018},
  organization={PMLR}
}

@article{lecun2002gradient,
  title={Gradient-based learning applied to document recognition},
  author={LeCun, Y. and Bottou, L. and Bengio, Y. and Haffner, P.},
  journal={Proc. IEEE},
  volume={86},
  number={11},
  pages={2278--2324},
  year={1998},
  publisher={Ieee}
}

@article{taylor1918dissipation,
  title={On the dissipation of eddies},
  author={Taylor, G. I.},
  journal={Meteorology, Oceanography and Turbulent Flow},
  pages={96--101},
  year={1918},
  publisher={Cambridge University Press}
}

@article{andreu2020effect,
  title={Effect of transverse gust velocity profiles},
  author={Andreu-Angulo, I. and abinsky, H. and Biler, H. and Sedky, G. and Jones, A. R.},
  journal={AIAA J.},
  volume={58},
  number={12},
  pages={5123--5133},
  year={2020},
  publisher={American Institute of Aeronautics and Astronautics}
}

@article{smith2024cyclic,
  title={A cyclic perspective on transient gust encounters through the lens of persistent homology},
  author={Smith, L. and Fukami, K. and Sedky, G. and Jones, A. and Taira, K.},
  journal={J. Fluid Mech.},
  volume={980},
  pages={A18},
  year={2024},
  publisher={Cambridge University Press}
}

@article{shannon1948mathematical,
  title={A mathematical theory of communication},
  author={Shannon, C. E.},
  journal={Bell Syst. Tech. J.},
  volume={27},
  number={3},
  pages={379--423},
  year={1948},
  publisher={Nokia Bell Labs}
}

@article{fukami2026information,
  title={Information-theoretic machine learning for time-varying mode decomposition of separated aerodynamic flows},
  author={Fukami, K. and Araki, R.},
  journal={AIAA J.},
  volume={64},
  number={2},
  pages={605--613},
  year={2026},
  publisher={American Institute of Aeronautics and Astronautics}
}

@article{fujino2023hierarchy,
  title={Hierarchy of coherent vortices in turbulence behind a cylinder},
  author={Fujino, J. and Motoori, Y. and Goto, S.},
  journal={J. Fluid Mech.},
  volume={975},
  pages={A13},
  year={2023},
  publisher={Cambridge University Press}
}

@article{kingma2014adam,
  title={Adam: A method for stochastic optimization},
  author={Kingma, D. P.},
  journal={\rm{arXiv:1412.6980}},
  year={2014}
}

@article{lumley1967structure,
  title={The structure of inhomogeneous turbulent flows},
  author={Lumley, J. L.},
  journal={Atmos. Turbul. Radio Wave Propag.},
  pages={166--178},
  year={1967},
  publisher={Nauka}
}

@article{linot2025extracting,
  title={Extracting dominant dynamics about unsteady base flows},
  author={Linot, A. and Lopez-Doriga, B. and Zhong, Y. and Taira, K.},
  journal={Fluid Dyn. Res.},
  volume = {57},
  number = {3},
  pages = {031401},
  year={2025}
}

@article{cremades2025classically,
  title={Classically studied coherent structures only paint a partial picture of wall-bounded turbulence},
  author={Cremades, A. and Hoyas, S. and Vinuesa, R.},
  journal={Nat. Commun.},
  volume={16},
  number={1},
  pages={10189},
  year={2025},
  publisher={Nature Publishing Group UK London}
}

@article{fukami2023grasping,
  title={Grasping extreme aerodynamics on a low-dimensional manifold},
  author={Fukami, K. and Taira, K.},
  journal={Nat. Commun.},
  volume={14},
  number={1},
  pages={6480},
  year={2023},
  publisher={Nature Publishing Group UK London}
}

@article{fukami2024data,
  title={Data-driven transient lift attenuation for extreme vortex gust--airfoil interactions},
  author={Fukami, K. and Nakao, H. and Taira, K.},
  journal={J. Fluid Mech.},
  volume={992},
  pages={A17},
  year={2024},
  publisher={Cambridge University Press}
}

@article{biler2021experimental,
  title={Experimental investigation of transverse and vortex gust encounters at low {R}eynolds numbers},
  author={Biler, H. and Sedky, G. and Jones, A. R. and Saritas, M. and Cetiner, O.},
  journal={AIAA J.},
  volume={59},
  number={3},
  pages={786--799},
  year={2021},
  publisher={American Institute of Aeronautics and Astronautics}
}

@article{brunton2020machine,
  title={Machine learning for fluid mechanics},
  author={Brunton, S. L. and Noack, B. R. and Koumoutsakos, P.},
  journal={Annu. Rev. Fluid Mech.},
  volume={52},
  number={1},
  pages={477--508},
  year={2020},
  publisher={Annual Reviews}
}

@article{taira2025machine,
  title={Machine learning in fluid dynamics: A critical assessment},
  author={Taira, K. and Rigas, G. and Fukami, K.},
  journal={Phys. Rev. Fluids},
  volume={10},
  number={9},
  pages={090701},
  year={2025},
  publisher={APS}
}

@article{fukami2025extreme,
  title={Extreme vortex-gust airfoil interactions at {R}eynolds number 5000},
  author={Fukami, K. and Smith, L. and Taira, K.},
  journal={Phys. Rev. Fluids},
  volume={10},
  number={8},
  pages={084703},
  year={2025},
  publisher={APS}
}

@book{davidson2015turbulence,
  title={Turbulence: an introduction for scientists and engineers},
  author={Davidson, P.},
  year={2015},
  publisher={Oxford {U}niversity {P}ress}
}

@article{ooi1999study,
  title={A study of the evolution and characteristics of the invariants of the velocity-gradient tensor in isotropic turbulence},
  author={Ooi, A. and Martin, J. and Soria, J. and Chong, M. S.},
  journal={J. Fluid Mech.},
  volume={381},
  pages={141--174},
  year={1999},
  publisher={Cambridge University Press}
}

@article{ashtiani2025data,
  title={Data-driven time-dependent bases for turbulent airfoil wake-extreme vortex gust interactions},
  author={Zamani Ashtiani, S. and Fukami, K.},
  journal={AIAA J. {\rm in press}},
  year={2026}
}

@article{morimoto2021convolutional,
  title={Convolutional neural networks for fluid flow analysis: toward effective metamodeling and low dimensionalization},
  author={Morimoto, M. and Fukami, K. and Zhang, K. and Nair, A. G. and Fukagata, K.},
  journal={Theor. Comput. Fluid Dyn.},
  volume={35},
  number={5},
  pages={633--658},
  year={2021},
  publisher={Springer}
}

@article{hansen1993use,
  title={The use of the {L}-curve in the regularization of discrete ill-posed problems},
  author={Hansen, P. C. and O’Leary, D. P.},
  journal={SIAM J. Sci. Comput.},
  volume={14},
  number={6},
  pages={1487--1503},
  year={1993},
  publisher={SIAM}
}

@article{jones2022physics,
  title={Physics and modeling of large flow disturbances: discrete gust encounters for modern air vehicles},
  author={Jones, A. R. and Cetiner, O. and Smith, M. J.},
  journal={Annu. Rev. Fluid Mech.},
  volume={54},
  number={1},
  pages={469--493},
  year={2022},
  publisher={Annual Reviews}
}

@article{taira2026extreme,
  title={Extreme aerodynamics: {A} data-driven perspective},
  author={Taira, K.},
  journal={Phys. Rev. Fluids},
  volume={11},
  number={1},
  pages={014702},
  year={2026},
  publisher={APS}
}

@article{goto2017hierarchy,
  title={Hierarchy of antiparallel vortex tubes in spatially periodic turbulence at high {R}eynolds numbers},
  author={Goto, S. and Saito, Y. and Kawahara, G.},
  journal={Phys. Rev. Fluids},
  volume={2},
  number={6},
  pages={064603},
  year={2017},
  publisher={APS}
}

@article{fukami2024data_pi,
  title={Data-driven nonlinear turbulent flow scaling with Buckingham Pi variables},
  author={Fukami, K. and Goto, S. and Taira, K.},
  journal={J. Fluid Mech.},
  volume={984},
  pages={R4},
  year={2024},
  publisher={Cambridge University Press}
}

@article{cremades2025additive,
  title={Additive-feature-attribution methods: {A} review on explainable artificial intelligence for fluid dynamics and heat transfer},
  author={Cremades, A. and Hoyas, S. and Vinuesa, R.},
  journal={Int. J. Heat Fluid Flow},
  volume={112},
  pages={109662},
  year={2025},
  publisher={Elsevier}
}

@article{ribeiro2023laminar,
  title={Laminar post-stall wakes of tapered swept wings},
  author={Ribeiro, J. M. and Neal, J. and Burtsev, A. and Amitay, M. and Theofilis, V. and Taira, K.},
  journal={J. Fluid Mech.},
  volume={976},
  pages={A6},
  year={2023},
  publisher={Cambridge University Press}
}

@article{odaka2026vorticala,
  title={Vortical similarities across laminar and turbulent extreme gust encounters},
  author={Odaka, H. and Lopez-Doriga, B. and Taira, K.},
  journal={J. Fluid Mech.},
  volume={1031},
  pages={R3},
  year={2026},
  publisher={Cambridge University Press}
}

@article{Koshikawa_Araki_Liu_Fukami_2026, title={Convolutional causal learning for aerodynamic flows}, volume={1037}, DOI={10.1017/jfm.2026.11699}, journal={J. Fluid Mech.}, author={Koshikawa, R. and Araki, R. and Liu, Q. and Fukami, K.}, year={2026}, pages={A6}}

@inproceedings{hornung1989vorticity,
  title={Vorticity generation and transport},
  author={Hornung, H.},
  booktitle={Tenth Australasian Fluid Mechanics Conference},
  pages={KS3.1--KS3.7},
  year={1989},
  organization={University of Melbourne}
}

@book{wu2012vorticity,
  title={Vorticity and Vortex Dynamics},
  author={Wu, J.-Z. and Ma, H.-Y. and Zhou, M.-D.},
  year={2012},
  publisher={Springer},
  address={Berlin, Heidelberg}
}

@article{cremades2026assessment,
  title={Assessment of non-intrusive sensing in wall-bounded turbulence through explainable deep learning},
  author={Cremades, A. and Freibergs, R. and Hoyas, S. and Ianiro, A. and Discetti, S. and Vinuesa, R.},
  journal={J. Fluid Mech.},
  volume={1028},
  pages={A19},
  year={2026},
  publisher={Cambridge {U}niversity {P}ress}
}

@article{barnes2018clockwise,
  title={Clockwise vortical-gust/airfoil interactions at a transitional {R}eynolds number},
  author={Barnes, C. J. and Visbal, M. R.},
  journal={AIAA J.},
  volume={56},
  number={10},
  pages={3863--3874},
  year={2018},
  publisher={American Institute of Aeronautics and Astronautics}
}

@article{barnes2018counterclockwise,
  title={Counterclockwise vortical-gust/airfoil interactions at a transitional {R}eynolds number},
  author={Barnes, C. J. and Visbal, M. R.},
  journal={AIAA J.},
  volume={56},
  number={7},
  pages={2540--2552},
  year={2018},
  publisher={American Institute of Aeronautics and Astronautics}
}

@article{ribeiro2024triglobal,
  title={Triglobal resolvent-analysis-based control of separated flows around low-aspect-ratio wings},
  author={Marques Ribeiro, J. H. and Taira, K.},
  journal={J. Fluid Mech.},
  volume={995},
  pages={A13},
  year={2024},
  publisher={Cambridge University Press}
}

@article{fukami2025observable,
  title={Observable-augmented manifold learning for multi-source turbulent flow data},
  author={Fukami, K. and Taira, K.},
  journal={J. Fluid Mech.},
  volume={1010},
  pages={R4},
  year={2025},
  publisher={Cambridge University Press}
}

@article{lozano2022information,
  title={Information-theoretic formulation of dynamical systems: causality, modeling, and control},
  author={Lozano-Dur{\'a}n, A. and Arranz, G.},
  journal={Phys. Rev. Res.},
  volume={4},
  number={2},
  pages={023195},
  year={2022},
  publisher={APS}
}

@article{lopez2026effect,
  title={On the effect of airfoil geometry on extreme vortex-gust encounters},
  author={Lopez-Doriga, B. and Jones, A. R. M. and Taira, K.},
  journal={J. Fluid Mech.},
  volume={1037},
  pages={A61},
  year={2026},
  publisher={Cambridge {U}niversity {P}ress}
}

@article{fukami2025compact,
  title={Compact representation of transonic airfoil buffet flows with observable-augmented machine learning},
  author={Fukami, K. and Iwatani, Y. and Maejima, S. and Asada, H. and Kawai, S.},
  journal={J. Fluid Mech.},
  volume={1021},
  pages={A39},
  year={2025},
  publisher={Cambridge University Press}
}

@article{koshikawa2027AISIN,
  title={Extracting the causal relationship of flow around an automobile with information-theoretic machine learning},
  author={Koshikawa, R. and Kishimoto, M. and Fukami, K.},
  journal={AIAA paper},
  year={2027},
}

@article{ham2004energy,
  title={Energy conservation in collocated discretization schemes on unstructured meshes},
  author={Ham, F. and Iaccarino, G.},
  journal={Annu. Res. Briefs},
  volume={2004},
  number={3-14},
  pages={118},
  year={2004}
}

@article{ham2006accurate,
  title={Accurate and stable finite volume operators for unstructured flow solvers},
  author={Ham, F. and Mattsson, K. and Iaccarino, G.},
  journal={Annu. Res. Briefs},
  volume={243},
  year={2006},
  publisher={Center for Turbulence Research, NASA Ames/Stanford University Stanford}
}

@article{fukami2020assessment,
  title={Assessment of supervised machine learning methods for fluid flows: K. Fukami et al.},
  author={Fukami, K. and Fukagata, K. and Taira, K.},
  journal={Theor. Comput. Fluid Dyn.},
  volume={34},
  number={4},
  pages={497--519},
  year={2020},
  publisher={Springer}
}

@article{mckeon2010critical,
  title={A critical-layer framework for turbulent pipe flow},
  author={McKeon, B. J. and Sharma, A. S.},
  journal={J. Fluid Mech.},
  volume={658},
  pages={336--382},
  year={2010},
  publisher={Cambridge University Press}
}

@article{odaka2026extremeb,
  title={Extreme vortex gust encounters by a square wing},
  author={Odaka, H. and Smith, L. and Taira, K.},
  journal={J. Fluid Mech.},
  volume={1037},
  pages={A67},
  year={2026},
  publisher={Cambridge University Press}
}

@inproceedings{hunt1988eddies,
  title={Eddies, streams, and convergence zones in turbulent flows},
  author={Hunt, J. C. R. and Wray, A. A. and Moin, P.},
  booktitle={Studying Turbulence Using Numerical Simulation Databases - II. Proceedings of the 1988 Summer Program},
  editor={Moin, P. and Reynolds, W. C. and Kim, J.},
  publisher={Stanford University},
  pages={193--208},
  year={1988}
}

@article{pearl2009causal,
  title={Causal inference in statistics: an overview},
  author={Pearl, J.},
  journal={Stat. Surv.},
  volume={3},
  pages={95--146},
  year={2009}
}

\end{document}